\documentclass[aps, prx, 10pt, twocolumn, floatfix, superscriptaddress, longbibliography]{revtex4-1}
\usepackage{amssymb,amsmath,amstext}
\usepackage{graphicx}
\usepackage{color}
\usepackage[utf8]{inputenc}
\usepackage{bbold}
\usepackage{bbm}
\usepackage{latexsym}
\usepackage[colorlinks=true,citecolor=blue,linkcolor=magenta]{hyperref}
\usepackage{enumerate}

\begin{document}

\title{Sampling diverse near-optimal solutions via algorithmic quantum annealing}

\author{Masoud Mohseni}
\affiliation{Google Quantum AI, Venice, CA 90291}
\affiliation{LSIP, Hewlett Packard Labs, Milpitas, California, US}

\author{Marek M. Rams}
\affiliation{Jagiellonian University, Institute of Theoretical Physics, \L{}ojasiewicza 11, 30-348 Krak\'{o}w, Poland}

\author{Sergei V. Isakov}
\affiliation{Google Quantum AI, Zurich, Switzerland}

\author{Daniel Eppens}
\affiliation{Google Quantum AI, Venice, California 90291}

\author{Susanne Pielawa}
\affiliation{Google, Munich, Germany}

\author{Johan Strumpfer}
\affiliation{Google, Mountain View, California 94043}

\author{Sergio Boixo}
\affiliation{Google Quantum AI, Venice, CA 90291}

\author{Hartmut Neven}
\affiliation{Google Quantum AI, Venice, CA 90291}

\begin{abstract}
Sampling a diverse set of high-quality solutions for hard optimization problems is of great practical relevance in many scientific disciplines and applications, such as artificial intelligence and operations research. One of the main open problems is the lack of ergodicity, or mode collapse, for typical stochastic solvers based on Monte Carlo techniques leading to poor generalization or lack of robustness to uncertainties. Currently, there is no universal metric to quantify such performance deficiencies across various solvers. Here, we introduce a new diversity measure for quantifying the number of independent approximate solutions for NP-hard optimization problems. Among others, it allows benchmarking solver performance by a required time-to-diversity~(TTD), a generalization of often used time-to-solution~(TTS). We illustrate this metric by comparing the sampling power of various quantum annealing strategies. In particular, we show that the inhomogeneous quantum annealing schedules can redistribute and suppress the emergence of topological defects by controlling space-time separated critical fronts, leading to an advantage over standard quantum annealing schedules with respect to both TTS and TTD for finding rare solutions. Using path-integral Monte Carlo simulations for up to 1600 qubits, we demonstrate that nonequilibrium driving of quantum fluctuations, guided by efficient approximate tensor network contractions, can significantly reduce the fraction of hard instances for random frustrated 2D spin-glasses with local fields. Specifically, we observe that by creating a class of algorithmic quantum phase transitions, the diversity of solutions can be enhanced by up to~$40\%$ with the fraction of hard-to-sample instances reducing by more than~$25\%$.
\end{abstract}

\maketitle

Sampling diverse solutions of combinatorial problems pose a significant difficulty due to the exponential explosion of the configuration space. This problem can be reformulated as finding independent low-energy states of spin-glass systems~\cite{mezard_information_2009}. Sampling is at the core of discrete optimization, robust optimization, and counting problems which are \#P-complete~\cite{moore_nature_2011}. Unbiased or fair sampling over discrete spaces is also one of the computational bottlenecks in machine learning tasks, including training and inference tasks in structured probabilistic models~\cite{goodfellow_deep_2016} and energy-based models~\cite{lecun_tutorial_2006}. In particular, approximating partition function or evaluating the marginal distributions for random Markov fields, Boltzmann machines, Hopfield models, or Bayesian networks are yet intractable over general graph topologies for high-dimensional data~\cite{goodfellow_deep_2016}.

Historically, there have been many different proposals to measure diversity in various fields, including statistics, biology, economy, and computer science. However, there is no universal measure to quantify the diversity of the ``types" of entities or elements in a population~\cite{page_diversity_2011}. Additionally, in the context of discrete optimization the notion of the types of solutions is not well-defined within the configuration space. Measures of variations such as the standard deviation are typically defined over a single parameter or attribute (e.g., residual energy), but they do not address diversity in the types of entities~\cite{page_diversity_2011}. Such diversity can be captured by generalized entropy measures encompassing the Simpson diversity index~\cite{simpson_measurement_1949} and Renyi or Shannon entropy~\cite{spellerberg_tribute_2003} as special cases. Entropy measures, however, do not have a build-in notation of distance among the types of solutions. Thus, as we show here, one has to introduce an acceptable metric to estimate the number of independent (pure) states within the desired approximation ratio.

Traditionally, certain measures such as Weitzmann diversity~\cite{weitzman_diversity_1992}, which defines the distance as the minimal number of links that connect two elements or solutions in treelike graphs, have been widely used in ecology and economy. However, constructing meaningful treelike structures is hard within configuration space, and these measures could lead to overestimating the diversity in our case. This difficulty in estimating diversity is due to exponentially dense sets of closely related solutions in various basins of attractions (pure states) that could form a hierarchical structure~\cite{mezard_information_2009}, or the existence of a few extremely eccentric solutions. The challenges of local stochastic solvers, such as parallel tempering, for a fair sampling of the degenerate ground states of Ising models, have been recently studied in Ref.~\cite{zhu_fair_2019}.

In operations research, the notion of diversity prominently appears in robust optimization and black-box optimization where the true cost function is not known, could be very noisy, or hard to evaluate~\cite{beyer_robust_2007}. In the context of multiobjective optimization, the diversity indicates the richness of states at the Pareto fronts that could be more robust to uncertainties. The fitness landscape of combinatorial optimization has been studied via numerical estimation of the number of basins of attraction and their neighborhoods~\cite{garnier_efficiency_2001}. However, such empirical metrics did not account for the quality of the local optima and were limited to simple local stochastic search algorithms that are based on the steepest descent and random restarts.

Diversity could be an essential feature in the construction of certain heuristic algorithms. For example, diversity is a key hyperparameter in genetic algorithms, population dynamics, ant colony optimization, particle swarm optimization, and evolutionary optimization~\cite{pugh_quality_2016}. In particular, the notation of \textit{quality diversity} has been introduced in the context of open-ended evolutionary optimization to capture high-performing solutions over certain phenotypic feature space~\cite{pugh_quality_2016}. The quality diversity has applications in adaptive robotic systems that could be inherently equipped with a diverse set of high-performing agents or policies to effectively cope with uncertainties, damages, and changing environmental conditions~\cite{cully_robots_2015}. This is in contrast with traditional machine learning methods that produce little diversity and typically fail if they are exposed to inputs slightly outside of a narrow scope defined by the training data~\cite{goodfellow_deep_2016}.

\begin{figure}[t]
\begin{centering}
\includegraphics[width=\columnwidth]{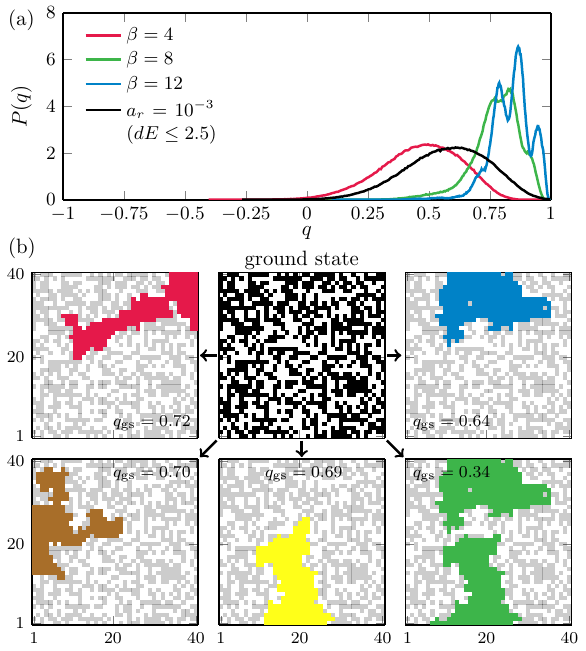}
\par\end{centering}
\caption{{\bf Diversity of the low-energy spectrum for the random Ising model on a 2D lattice.}
In this plot, we consider a single random instance of an Ising Hamiltonian in Eq.~\eqref{eq:HP} with $1600$ spins defined on a square lattice with nearest-neighbor couplings $J_{ij} \in [-1, 1] $ and local fields $h_i \in [-0.1, 0.1]$ drawn from uniform distributions.  In panel (a), we show the distributions $P(q)$ of overlaps $q$ between pairs of states independently sampled from the Gibbs distribution for various temperatures. The black curve shows $P(q)$ if all the low-energy states would be equally probable within approximation ratio $a_r=0.001$. This illustrates that $P(q)$ distributions are inadequate to capture the diversity of low-energy states. These distributions are essentially featureless in the low-energy spectrum of interests, and they cannot distinguish distinct droplet excitations of similar volume leading to indistinguishable values of $q$. Those are shown in panel (b), where the top-middle panel indicates the ground state configuration of the considered instance. Other panels show some of the other distinct low-energy configurations within the approximation ratio of $a_r=0.001$. Those states are obtained by starting in the ground state and then flipping the groups of neighboring spins, which are indicated by colorful droplets, leading to minimal excitations above the ground state within the target approximation ratio and informing on the geometry of the low-energy manifold.
\label{fig:1}}
\end{figure}

For spin-glass systems, the notion of distance or dissimilarity for solutions is usually captured by the Parisi's order parameter~\cite{parisi_order_1983} that can be approximated by the probability distribution function $P(q)$, where $q_{ab} := \frac{1}{N} \sum_{i=1}^N s_i^{(a)} s_i^{(b)}$ is the overlap of two randomly sampled pairs of spin configurations $s^{(a)}$ and $s^{(b)}$, where $s_i=\pm1$. However, this metric provides a simple one-dimensional projection of the solution space and thus cannot be used to quantify the diversity of independent solutions, since by construction it doesn't distinguish the contributions from different types of solutions of the spin-glass pure states in effectively high-dimensional space, see Fig.~\ref{fig:1}.

The geometry of near-optimal solutions, and the hardness of sampling such low-energy states, can be captured by the so-called \emph{overlap gap property} (OGP)~\cite{gamarnik_finding_2018,gamarnik_overlap_2021}, where clusters of solutions become shattered or disjoint near a computational phase transition. OGP is present when Hamming distances between any pair of solutions within some approximation ratio, is either smaller than $\nu_1$ or greater than $\nu_2$ for some fixed $\nu_1 < \nu_2$. A useful definition of diversity should be able to provide an estimate of the number of independent solutions when OGP holds.

In this work, we introduce such a measure of diversity by quantifying the maximum number of likely independent solutions that all satisfy a given cost function(s) for a desired high precision and are unlikely to be related to each other via a set of local updates. We also introduce a diversity ratio as the fraction of such high-quality, independent solutions that a given solver can find, in a given timescale, normalized by the absolute value of diversity. We show that our diversity ratio is separate from the approximation ratio and can be used as a new probe to quantify the performance of several quantum and classical heuristic optimization algorithms.

In particular, we compare algorithmic quantum annealing strategies (with spatial and temporal inhomogeneities) versus standard adiabatic quantum computing.  We use efficient approximate tensor-network contractions as a preprocessing step to create multiple critical fronts which could effectively control (suppress and/or redistribute) the presence or location of topological defects in many different ways, minimizing the energy cost of solutions or increase their diversity. We numerically simulate the quantum annealing protocol for sampling rare solutions in a quasi-1D transverse Ising models of up to 512 qubits using matrix product state (MPS)~\cite{verstraete_matrix_2008,schollwock_density-matrix_2011} techniques. We use quantum Monte Carlo techniques~\cite{isakov_understanding_2016} to simulate multicritical annealing fronts of the quantum spin glass dynamics for 2D systems of up to 1600 qubits. Using inhomogeneous annealing schedules, we observe that the diversity of solutions can be enhanced by about $40\%$ within a timescale of $10^{9}$ sweeps. Moreover, the fraction of the hard instances can be reduced by more than $25\%$ for capturing at least a diversity of $50\%$ of high-quality solutions on each random instance.

\section{Diversity measure}

In this article, we introduce a notion of diversity that is an inherent characteristic of the spin-glass configuration representing problem instances and not an attribute of any classical or quantum solvers employed to tackle the problem. We focus on spin-glass Ising Hamiltonians with quadratic interactions
\begin{equation}
H_{P} = \sum_{i<j} J_{i,j} s_i  s_j + \sum_{i=1}^N h_{i} s_i,
\label{eq:HP}
\end{equation}
where $J_{i,j}$ and $h_{i}=J_{i,i}$ are coupling interactions and local fields encoding the problem specification for each instance containing $N$ binary variables $s_i = \pm 1$. However, our diversity measure definition can be extended to other generalized models over discrete spaces with nonbinary variables and/or higher order interactions, such as factor graphs or k-SAT problems with $k \geq 3$~\cite{mezard_information_2009} or modern Hopfield networks~\cite{krotov_hopfield_2016}.

The widely used measure to benchmark various heuristic  solvers is time-to-solution (TTS)~\cite{ronnow_defining_2014}, or more generally time-to-approximation-ratio;
that is the time needed to find (with a given certainty) at least a solution whose energy is within the desired low-energy manifold, see Appendix~\ref{sec:app1}. The targeted manifold's width is typically taken as a fraction, i.e., \emph{approximation ratio} $a_r$, of the total energy bandwidth. However, one is often interested in having a protocol that not only gives excellent residual energies but can also effectively sample from a variety of {\it distinct} solutions. To quantify this, below we introduce {\it time-to-diversity} (TTD).

Let us first consider all the low-energy states within a given approximation ratio above the ground state. The goal is to divide those states into classes, or \textit{types}, such that the states belonging to different classes differ significantly. We define such types or clusters of solutions in the configuration space by dividing the low-energy spectrum into distinct basins of attraction as far as they are \textit{distant} from all other such clusters according to an acceptable metric,
\begin{equation}
    d({s}^{(a)},{s}^{(b)}) \ge R N.
    \label{eq:dindependence}
\end{equation}
Here, ${s}^{(a)}$ and ${s}^{(b)}$ are any two configurations belonging to the set, $R \in [0,1]$ is the normalized distance threshold, and we take $d({s}^{(a)},{s}^{(b)})$ as a Hamming distance (for low-dimensional problems, one may consider its refinement, which we discuss below). In the thermodynamics limit, these basins of attractions could correspond to a subset of the pure spin-glass states that are mutually distant according to such metric. Spin glass pure states can be characterized within one-step replica symmetry breaking (1RSB) cavity method~\cite{mezard_information_2009}.

Formally, diversity measure can be evaluated as follows: (i) We estimate the set of low energy states for a  given approximation ratio $a_r$. (ii) We build an undirected graph, $G=(V,E)$, where each vertex or node, $V$, correspond to a low energy state ${s}$, their edges $E$  has weights correspond to their mutual Hamming distance $d({s}^{(a)},{s}^{(b)})$ and we ignore all the edges with weight larger than $RN$. (iii) The diversity measure, $D$, becomes the cardinality of the maximal independent set~\cite{moore_nature_2011} over this graph. That is all the vertices that have mutual Hamming distance larger than $RN$ (for a system of $N$ spins); e.g., for $R=1/8$ this number will likely correspond to independent low-energy states that belong to different pure states of the original spin-glass encoding the problem (by having larger $R$  we can increase our confidence on the independence of these low energy states). We show an example of such a maximal independent set in Fig.~\ref{fig:1}(b), where $D=6$ for $R=1/8$. (iv) Diversity ratio, $d_r$, is the total number of independent low energy states that one can find using a given solver (possibly limited to some total computational time) for a given approximation ratio, $D_{\rm solver}$, over the absolute diversity at the same approximation ratio; that is $d_r=D_{\rm solver}/D$.

We note the threshold value of the Hamming distance characterized by $R$ in Eq.~\eqref{eq:dindependence} can be selected according to the parameters introduced within the OGP ~\cite{gamarnik_overlap_2021}. That is, whenever the OGP holds for a given problem (when mutual distance between any two solutions is smaller than $\nu_1$ or larger than $\nu_2$, where $\nu_1 < \nu_2$) the relevant range of values for $RN$ would be inside the gap; i.e., between $\nu_1$ and $\nu_2$. That implies that we are effectively counting the number of independent solutions since their distances will be greater than $\nu_1$.

The evaluation of diversity measure $D$ can be also described by maximum clique problem~\cite{bomze_maximum_1999} on the complement graph of $G$ in which any two low-energy vertices with mutual Hamming distance of $d({s}^{(a)},{s}^{(b)}) \ge R N$ become adjacent. The maximum clique problem is known to be NP-complete. However, several heuristic algorithms have been developed, such as the approach by Balaji, Swaminathan, and Kannan~\cite{balaji_simple_2010} which runs in $O(n^{2})$ time. In this article, we use a greedy procedure to approximate it and identify the seeds for basins of attractions. We elucidate more on this in the context of our examples.

The {\it time-to-diversity-ratio} for a given solver can be now evaluated by the time (including restarts and repetitions) needed to find low-energy configurations belonging to at least $d_r  D$ basins of attraction seeded by the solutions of the above max-clique problem, where $d_r$ is the desired diversity ratio. We assign a configuration to a given basin if it is closer to its seed than any other seed.

The above-described procedure poses some challenges in itself -- which is inevitable for spin-glass problems. In practice, the total diversity, $D$, can be calculated only within our best knowledge of the solution's space. However, that limitation is also true for other well-known metrics to quantify the optimality of solutions, such as approximate ratio. In the absence of any provable bounds for best solutions, one can combine the results from a portfolio of solvers, calculate the total diversity, and evaluate $d_r$ and TTD for individual solvers. The results can be improved iteratively by updating the baseline benchmarks once new distinct classes of solutions get identified for a specific problem. This procedure provides a natural platform to compare different solvers. In the next sections, we test our diversity measure to quantify the computational power of inhomogeneous quantum annealing strategies~\cite{mohseni_engineering_2018} against standard adiabatic quantum computing.

Finally, it is worth to further discuss the measure of distance between two spin configurations, $d({s}^{(a)}, {s}^{(b)})$, that appears in Eq.~\eqref{eq:dindependence}. The most simple choice is a Hamming distance between ${s}^{(a)}$ and ${s}^{(b)}$, i.e., the number of spins where the two configurations differ. However, as we argue here, that choice has to be further refined. A problem with this measure is that certain solutions that are far from each other in term of the Hamming distance could still be connected. In such cases, there could be no significant energy barriers to local stochastic search algorithms for navigating among them; e.g., variations in many small clusters of variables could still add to a large effective Hamming distance. A natural choice for refinement of Hamming distance, in particular for low-dimensional systems, is to look at the number of spins in the largest {\it singly connected} cluster of spins (according to the adjacency matrix of $J_{i,j}$) where the two configurations ${s}^{(a)}$ and ${s}^{(b)}$ differ.

This choice is also motivated by droplet excitations in spin-glass systems~\cite{fisher_equilibrium_1988}. This singly connected Hamming distance captures the situations that relatively large clusters of variables conspire together, due to intrinsic interplay of disorders and/or frustration. This is at core of computational complexity of random k-SAT problem where a large set of frozen variables, or the backbone~\cite{mezard_information_2009} could emerge deep into a rigidity region near a computational phase transition~\cite{moore_nature_2011}. The mixing time of many heuristics with local updates such as Markov chain Monte Carlo (MCMC) grow exponentially with the size of such droplets or backbones since the state of Monte Carlo sampler is essentially pinned to a single basin of attraction for sufficiently low temperatures. Indeed, Houdayer or Iso-energetic cluster moves (ICM) are designed with the hope to overcome such shortcoming of local stochastic updates~\cite{houdayer_cluster_2001, zhu_efficient_2015}. For the aforementioned reasons, in this work, we use such a refined singly connected Hamming distance.
We note that for a problem where variables reside on a fully-connected graph, the refined Hamming distance coincides with the standard Hamming distance.

\section{Overview of numerical examples}
In this article, we numerically study two sets of exemplary problems: (i) a quasi-one-dimensional (quasi-1D) setup with nonzero $J_{i,j}$ for $1 \le |i-j| \le r$, where the parameter $r$ sets the maximal range of interaction along the chain, and (ii) a two-dimensional (2D) $N=L \times L $ square lattice with nearest-neighbor interactions and local fields.

For the quasi-1D setup, we take $J_{i,j}$ as random uniformly distributed in $[-1, 1]$ with $r=3$ where each spin can be connected with up to six neighbors, see the inset of Fig.~\ref{fig:3}(a). We note that for $r=1$, one would recover a one-dimensional random Ising model. Transverse field Ising models in 1D have been exhaustively studied and well understood. They allow for various efficient approximation techniques, like the Strong-Disorder Renormalization Group approach~\cite{fisher_random_1992,fisher_critical_1995,  igloi_strong_2005}. These systems also allow for efficient exact simulation of quantum quenches, using the Jordan-Winger transformation to map the dynamics into a free-fermionic picture~\cite{dziarmaga_dynamics_2006, caneva_adiabatic_2007, rams_inhomogeneous_2016, mohseni_engineering_2018}. However, such a choice of $r=1$ prohibits frustrated classical ground states in the absence of quantum fluctuations. In this article, we consider quasi-1D systems for $r>1$, which could contain frustration.

For the 2D case, we also select random $J_{i,j}$ from the same random distribution in $[-1, 1]$, but we additionally include relatively weak local fields $J_{i,i}$ from a random uniform distribution in $[-0.1, 0.1]$. The addition of nonzero local fields precludes the polynomial-time exact ground state solver on a planar graph~\cite{khoshbakht_domain-wall_2018}. The 2D Ising model with local fields is computationally universal, and in principle can simulate the physics of all other higher dimensional spin-glass systems with polynomial embedding overhead~\cite{de_las_cuevas_simple_2016}.

This set of problems allows us to use efficient approximate numerical methods to get insight into their low-energy manifold as well as to numerically simulate quantum annealing protocols. Specifically, in Sec.~\ref{sec:base}, we briefly discuss a tensor-network based algorithm~\cite{rams_spin-glass_2021} that for intermediate-size quasi-2D random Ising problems in Eq.~\eqref{eq:HP} allows uncovering the manifold of low-energy solutions with great accuracy. We use the results of that procedure as a ground truth to calculate the diversity of low-energy solutions, as a reference to quantify the performance of quantum annealing, and to set up algorithmic annealing protocols with multiple spatio-temporal fronts. The latter are briefly motivated in Sec.~\ref{sec:inhomo}.

In Sec.~\ref{sec:quench} we study the performance of quantum annealing. In particular, we explore if the knowledge of rough droplet shapes can help to improve the sampling power of quantum annealing protocols via inhomogeneous driving schemes, as measured by TTS and TTD. To simulate the time evolution generated by the transverse field, in the quasi-1D case, we use MPS representation~\cite{verstraete_matrix_2008,schollwock_density-matrix_2011}. Here, it provides an effectively numerically exact method to simulate the quench dynamics: as the entanglement along the chain remains limited, the MPS ansatz with finite bond dimension provides a faithful representation of the state of the system during the ramp. For the 2D setup, we emulate quantum fluctuations using quantum Monte Carlo~\cite{isakov_understanding_2016} simulations.

\section{Ground truth diversity for low dimensional spin glass models}
\label{sec:base}
The considered classes of problems allow us to identify low-energy solution subspace for intermediate problem sizes with high confidence. To that end, we employ the tensor-network-based~\cite{verstraete_matrix_2008,schollwock_density-matrix_2011,biamonte_tensor_2017,orus_tensor_2019,ran_tensor_2020} approach of Ref.~\cite{rams_spin-glass_2021}. Below, we will briefly summarize it, while we refer to Ref.~\cite{rams_spin-glass_2021} for further details.

The method is based on representing the partition function of the classical spin model at small but finite temperature as a PEPS tensor network~\cite{nishino_two-dimensional_2001}. The partition function of the model and, more importantly, the marginal and conditional probabilities for any spin subconfiguration can be obtained by contracting the network. While the exact contraction of such a network would scale exponentially with the system tree-width, powerful strategies exist to {\it approximately} contract it for 2D geometries. In this work, we employ a boundary-MPS approach, see, e.g., Refs.~\cite{verstraete_matrix_2008, lubasch_unifying_2014}, which amounts to a transfer matrix method. In the exact simulation, the transfer matrices---and boundary vectors at which they act---would scale exponentially with the lattice width $L$. Here, they are efficiently represented using a compact one-dimensional structure of MPS. Such representation allows approximating the boundary-MPS systematically after each application of the transfer matrix, where each row of a 2D lattice corresponds to a transfer matrix. This is done while retaining the manageable size of the approximate boundary-MPS during the contraction of the network and calculation of conditional probabilities.

The approach of Ref.~\cite{rams_spin-glass_2021} combines the above-approximated tensor-network contraction technique with a branch-and-bound strategy to essentially map the low-energy spectrum. A sufficiently small temperature of the contracted distribution allows us to view and resolve the low-energy states. However, too low a temperature makes the approximate contraction numerically unstable. Successful execution of the algorithm requires simultaneously satisfying both those constraints, which puts a practical limitation on problem sizes studied in this article for the 2D case.

We scan the 2D system row after row to identify the most probable spin configurations, systematically building partial configurations and keeping track of those with the largest marginal probabilities. We further employ the locality of the interactions on the low-dimensional grid; namely, the energies and conditional probabilities for any region of the lattice depend only on the orientation of spins directly bordering this region. This locality is then used to identify equivalent partial configurations with respect to the division of the lattice into spins that have been considered at a given step of the branch-and-bound search versus those that have not yet considered or not attributed any values. Those equivalent configurations look identical from the point of view of not-yet considered part of the lattice and can be merged during the search. This merging reveals sub-configurations with the lowest conditional energies (conditioned on elements of the configuration that are connected with not-yet considered part of the lattice) and excitations above such partial ``ground states" encoded through spin-glass droplets. The systematic application of such a procedure leads to significant compression of the low-energy manifold, avoiding exhaustive Monte Carlo search, see, e.g., Refs.~\cite{houdayer_low-temperature_2004,zhu_fair_2019,barash_estimating_2019}, or targeting single low-energy excitations (droplets) in a planar problem without local fields~\cite{hartmann_large-scale_2002, khoshbakht_domain-wall_2018}. It also goes beyond counting the number of solutions of constraint satisfaction problems represented as tensor network contraction~\cite{garcia_sanchez_SAT_2012,biamonte_tensor_2015,kourtis_fast_2019}, or identifying the ground state via a combination of automatic differentiation with an {\it exact} contraction of tensor network with the topical algebra expressing the logarithm of the partition function in the zero-temperature limit~\cite{liu_tropical_2021}.

\begin{figure*}[t]
    \begin{centering}
    \includegraphics[width=2\columnwidth]{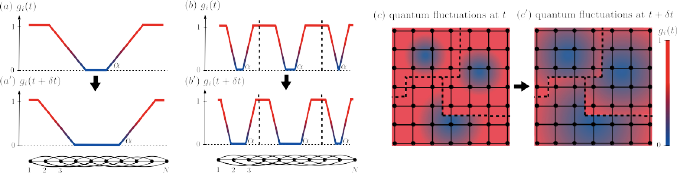}
    \par\end{centering}
    \caption{{\bf Schematic of time evolution and shape of multiple quantum critical fronts for disordered quasi-1D and 2D Ising models.} We show a few sample profiles for quantum fluctuations driven by spatio-temporal inhomogeneous external magnetic field $g_i(t)$ in Eq.~\eqref{eq:Ht}. This allows for spins (variables) that already experience the phase transition to influence the other variables in many different ways, essentially redistributing and suppressing defects, thus increasing the diversity of solutions.
    The value of the external magnetic field $g_i(t)$ is indicated with color-gradient changing between the initial (red) and final values (blue). Black dots indicate spins, and solid lines are the nonzero couplings $J_{i,j}$. We show a quasi-1D chain with a single cluster in panel (a) and multiple clusters in panel (b). The dashed lines indicate borders between clusters. We discuss the specific procedure we used to obtain them for numerical experiments in this work in Sec.~\ref{sec:quench}. In panel (c), we show time snapshots of the magnetic fields for a 2D lattice divided into three clusters.
    \label{fig:2}}
    \end{figure*}

As a first approximation, the number of large distinct droplet excitations can be related to the logarithm of diversity measure for systems with nonoverlapping droplets of sufficiently low energy.
In a more general setting, the diverse (independent) states of the low-energy manifold can be encoded through ground state configuration and a complicated hierarchical structure of droplet excitations on top of that ground state~\cite{rams_spin-glass_2021}. Those droplets indicate groups of spins that have to be flipped to jump from one local minimum to another one. An example of some large droplets above the ground state is shown in Fig.~\ref{fig:1}(b) for a 2D lattice.

We apply the above procedure both for quasi-1D and 2D setup. In the former, it is not particularly hard to extract all the low-energy states within the approximation ratio $a_r=0.0005$, which we employed in our examples (we later use this knowledge to calculate the probability of any such configuration following MPS time-dependent quench simulations).
Such a feat becomes impossible in the 2D setup as the total number of low-energy classical configurations (with $a_r=0.001$ used here) is enormous, particularly for the largest lattices of $N=40\times40$ spins that we consider. In the latter case, even a compressed description of the full low-energy manifold produced by such an algorithm  is prohibitive for $a_r$ of interests. We introduce the course-graining in the above merging procedure to overcome such limitations, discarding droplets with sizes below a few spins cut-off (which still can add up during consecutive merges). Fundamental limitations of that approach are ultimately related to the finite numerical precision and the accuracy of the approximate boundary-MPS, though we corroborate the convergence by repeating simulations for different temperatures and sizes of the boundary-MPS representation, see Appendix~\ref{sec:app2} for further numerical details.

To approximate (from below) the ground truth diversity $D$ and to identify the seeds of basins of attraction, we resort to a simple greedy algorithm. We start with the ground state configuration and iterate over the rest of identified states by an order that is indexed according to their residual energy above the ground state. The particular state becomes a seed of a new basin if its distance from each of the already identified seeds is larger than the desired normalized distance that we set at $R=1/8$.  Such a procedure is giving us an approximation for the true diversity measure $D$ for each instance.  We show the result of such a procedure for a particular disorder instance in Fig.~\ref{fig:1}(b), where we identify $D=6$ distinct attraction basins. Smaller droplets, that have not been shown in the plot, can be consequently flipped to further explore each basin of attraction.
We provide further information on identified $D$ for various setups and system sizes in Table~\ref{tab:s1}.

\begin{table}[b!]
\begin{tabular}{r|ccc}
& $q=20\%$ & $q=50\%$  & $q=80\%$ \\
\hline
\hline
quasi-1D $N=128$ & $1$ & $2$ & $2$ \\
quasi-1D $N=256$ & $3$ & $5$ & $8$ \\
quasi-1D $N=512$ & $13$ & $16$ & $24$ \\
\hline
2D $L=30$ & $4$ & $7$ & $11$ \\
2D $L=40$ & $4$ & $6$ & $12$ \\
\hline
\hline
\end{tabular}
\caption{{\bf Total diversity $D$ of the studied examples}. The results are for approximation ratio $a_r=0.0005$ for a quasi-1D setup, and $a_r=0.001$ for a 2D setup. We show quantiles: $20\%$, $50\%$ (median), and $80\%$ of 100 disorder instances. They follow from the approximate tensor-network-contraction-based branch-and-bound calculations to identify the low-energy states, followed by a greedy algorithm to identify independent configurations with normalized distance between any pair of configurations above $R=1/8$.
\label{tab:s1}}
    \end{table}

\section{Controlling inhomogeneous quantum phase transitions}
\label{sec:inhomo}
The approach overviewed in the previous section gives us reference solutions and their diversity for the considered set of problems.
We can now explore a diversity of solutions in quantum algorithms. In particular, we focus on a class of solvers employing quantum fluctuations induced by the transverse fields within a standard adiabatic quantum annealing paradigm. We also consider their generalization to quasi-adiabatic inhomogeneous quantum annealing and explore whether they may provide a way to improve TTS or TTD. Here, we briefly motivate this generalization.

\begin{figure}[t]
\begin{centering}
\includegraphics[width=0.95\columnwidth]{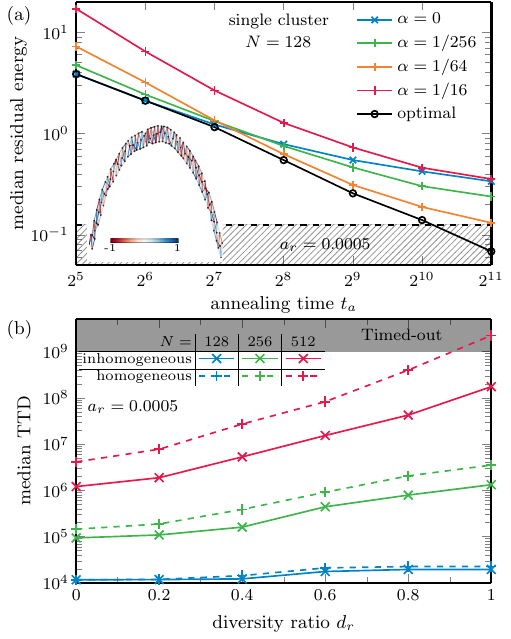}
\par\end{centering}
\caption{{\bf Inhomogeneous quenches for quasi-1D frustrated Ising model.}
    In panel (a), we show the reduction of the residual energy allowed by intermediate values of the inhomogeneous front's slope $\alpha$, comparing with the standard homogeneous protocol where $\alpha=0$. The latter visibly flatten out at long $t_a$, indicating slower-than-power-law dependence on the annealing time $t_a$.  While the best residual energy (here we plot median of 100 instances) is obtained for $\alpha=1/64$ at long times, we note that an optimal protocol is instance-dependent: black circles that represent the best results from the set of available $\alpha$'s give smaller energy than selecting the same single $\alpha$ for all instances. Here, we employ an inhomogeneous driving protocol with one cluster, as depicted in Fig.~\ref{fig:2}(a).
In panel (b), we focus on TTD and a protocol involving multiple fronts [see, Fig.~\ref{fig:2}(b)]. With increasing system size and targeted diversity ratio $d_r$, an inhomogeneous driving with a portfolio of $\alpha$'s allows outperforming the standard homogeneous approach. It becomes particularly relevant for some hard instances (see scattered plots in Fig.~\ref{fig:s1} for further evidence), outside of the regime of small-size effects where quick repetitions still turn out to optimize TTD; here, the results for each point are optimized with respect to $t_a$.
\vspace{-1pt}
\label{fig:3}}
\end{figure}

Inhomogeneous driving protocols as shortcuts to adiabaticity~\cite{guery-odelin_shortcuts_2019,del_campo_focus_2019} have been motivated by studies of the Kibble-Zurek mechanism~\cite{kibble_implications_1980, zurek_cosmological_1985}. The critical front is taking the system across a critical point one part after another; see Fig.~\ref{fig:2}. Light cones or causal zones that are forming can best explain the reduction of defects during a quantum quench as a part of the system that had crossed the critical point earlier can bias the part of the system crossing it later. That can happen if the spatial velocity of the inhomogeneous front is smaller than the velocity with which the information propagates in the system. This intuition applies both to classical~\cite{kibble_phase_1997,dziarmaga_symmetry_1999,zurek_causality_2009,del_campo_structural_2010,del_campo_inhomogeneous_2011} and quantum system~\cite{dziarmaga_dynamics_2010,*dziarmaga_adiabatic_2010,collura_critical_2010,*collura_nonlinear_2011,rams_inhomogeneous_2016,mohseni_engineering_2018,agarwal_fast_2018,susa_exponential_2018,susa_quantum_2018,gomez-ruiz_universal_2019,hartmann_quantum_2019,adame_inhomogeneous_2020,sinha_inhomogeneity_2020}. For quantum phase transitions, the examples include crossing the continuous critical points~\cite{dziarmaga_dynamics_2010,*dziarmaga_adiabatic_2010,collura_critical_2010,*collura_nonlinear_2011,gomez-ruiz_universal_2019}---including the case with long-range interactions~\cite{sinha_inhomogeneity_2020} or preparing the critical state itself~\cite{agarwal_fast_2018}, first-order transitions within mean-field treatment~\cite{susa_exponential_2018,susa_quantum_2018}, and unfreezing the Griffiths singularities for disordered systems~\cite{rams_inhomogeneous_2016,mohseni_engineering_2018}. The quantum case can be understood via opening the energy gap---or engineering the structure of the low-energy spectrum~\cite{mohseni_engineering_2018}. This casually induced quantum gap is dictated by the shape of the front and universal many-body properties of the critical point. In Fig.~\ref{fig:3}(a), we illustrate that such a mechanism can also be applied in frustrated systems, where we show that inhomogeneous driving can, in some cases, substantially decrease the residual energy in the quasi-1D setup with random interactions.

Multiple critical fronts~\cite{mohseni_engineering_2018}, which simultaneously take separate parts of the system through the transition, can considerably speed up the process, however, at the cost of creating defects between merging clusters. While detrimental at first sight, the multiple driving fronts provide a new mechanism to control the light cones influencing where the defects are most likely created. This mechanism could change the probability distribution of defects to have a high presence on the  desired places with low $J_{ij}$ minimizing the energy cost. This essentially allows one to target a low-energy manifold in a controllable manner by adjusting the shape and speed of critical fronts to create a diverse domain walls that have minimal cost. Here we apply such a mechanism to frustrated systems, both to reach the complicated, potentially degenerate, ground-state energies and as a strategy to explore the low-energy subspace capturing a diversity of near-optimal solutions.

The Hamiltonian of generalized transverse field Ising model with space-time separated critical fronts can be written as
\begin{equation}
\label{eq:Ht}
\hat H(t) = \hat H_{P}(t) - \sum_{i=1}^N g_i(t) \hat \sigma^x_i,
\end{equation}
with $\hat \sigma^z_i$ and $\hat \sigma^x_i$ being the standard Pauli operators for an $i$th spin. The Hamiltonian $\hat H_P(t) = \sum_{i<j} J_{i,j}(t) \hat  \sigma^z_i \hat  \sigma^z_j + \sum_{i=1}^N J_{i,i}(t) \hat \sigma^z_i$, which is diagonal in the computational basis, encodes the classical Ising Hamiltonian in Eq.~\eqref{eq:HP}. The transverse fields vary smoothly in time between $g_i(0) = 1$ for the initial $t=0$, and $g_i(t_a) = 0$ for the final or annealing time~$t_a$ (the unit of time is fixed by setting $\hbar=1$ and the amplitude of couplings $|J_{i,j}| \le 1$). We will use a protocol where the couplings are gradually switched-on in time, proportional to the switching-off of the transverse field. This entails introducing time-dependent couplings as $J_{i,j}(t) = (1 - g_i(t)/2 - g_j(t)/2) J_{i,j}$, so that at the initial time $\hat H(0) = - \sum_{i=1}^N g_i(t) \hat \sigma^x_i$, and the system is initialized in the ground state, being the equal-weights superposition of all classical configurations. At the final time, $\hat H(t_a)$ corresponds to the original problem~$H_P$.

As a possible strategy to improve the diversity of solutions, we make the fields $g_i(t)$ explicitly space-time dependent. To specify the driving protocol, we are going to divide the lattice into $M$ nonoverlapping clusters, see Fig.~\ref{fig:2}, and here we assume the transverse field of the form
\begin{equation}
\label{eq:general_front}
g_i(t)=\sum\limits_{k=1}^{M} f^{(k)}{\left(d^{(k)}_i - v^{(k)} t\right)},
\end{equation}
where $f^{(k)}$ is nonzero for spins inside $k$th cluster, and zero for spins that do not belong to the cluster, $d^{(k)}_i$ is a distance of $i$th lattice site from the center of the cluster and a function $f^{(k)}$ determines the shapes of the propagating fronts.

In this work, we use Euclidean distance on 1D and 2D lattice and mean position of the spins belonging to the cluster to specify its center. We assume time-independent velocity $v^{(k)}$, which depends on the maximal $d^{(k)}_i$ for spins belonging to the $k$th cluster, $d^{(k)}_{\mathrm{max}}$, and on the shape of the front encoded by $f^{(k)}$. Here, we take a linear front with the (possibly cluster-dependent) spatial slope $\alpha^{(k)}$,
\begin{equation}
\label{eq:gk}
f^{(k)}\left(d^{(k)}_i - v^{(k)} t\right) = [1 + \alpha^{(k)} (d^{(k)}_i - v^{(k)} t ) ]_{0,1},
\end{equation}
where we limit the possible values to lie between $0$ and $1$, introducing $[x]_{0,1} = \max(0, \min(1, x))$. Finally, the velocity $v^{(k)} = \frac{1+ \alpha^{(k)} d^{(k)}_{\mathrm{max}}}{\alpha^{(k)} t_a}$, where $t_a$ is the annealing time. Such a front is initialized at the center of the cluster and  propagates toward the cluster's boundaries, see Fig.~\ref{fig:2}, utilizing total available time $t_a$. For alternative constructions of the driving protocol with multiple critical fronts, see Ref.~\cite{mohseni_engineering_2018}.

In the limit of $\alpha^{(k)} \to 0$, we recover the standard homogeneous protocol with
$g_i(t) = 1 - \frac{t}{t_a}$; however, then the spatial velocity $v^{(k)} \to \infty$, limiting effective causal communication.  In the limit of $\alpha^{(k)} \to 1$, one recovers a one-spin-at-a-time protocol limiting the spatial extend of quantum fluctuations. This intuition allows us to expect optimal results for some intermediate values of the slope~$\alpha^{(k)}$.

\section{Constructing nonequilibrium quenches for enhanced diversity}
\label{sec:quench}
Finally, we can quantify the performance of annealing protocols for our set of problems and, in particular, explore the potential gains offered by an inhomogeneous driving strategy. We start in Fig.~\ref{fig:3} with an example of a connectivity graph forming quasi-1D chains. Figure~\ref{fig:3}(a) shows residual energies for relatively short chains of $N=128$ spins and inhomogeneous driving protocol where all system spins form a single cluster, corresponding to a driving protocol pictorially shown in Fig.~\ref{fig:2}(a). The latter driving strategy allows for a noticeable reduction of the excitation energy for longer annealing times, reducing the number of generated defects. However, the optimal slope of the inhomogeneous front turns out to be instance-dependent, which suggests using a portfolio of protocols with various slopes.

Gains in the residual energy (in particular for longer annealing times) do not have to directly translate to TTS, where the scenario involves multiple repetitions of the quench followed by measurement of the resulting classical configuration. We study the latter in Fig.~\ref{fig:3}(b), where we also consider inhomogeneous protocols driven within multiple clusters, see Fig.~\ref{fig:2}(b). As a proof of principle, here we set the borders of clusters to approximately correspond to the low-energy droplets outputted by the tensor-network branch-and-bound algorithm described in Sec.~\ref{sec:base}; see below for further details. We should stress here that rough droplets' boundaries are only needed here, and we do \textit{not} include information about the low-energy spin configurations. As a baseline study, we tested our protocol for uniform distribution of random clusters without observing noticeable gains. This confirms the intuition that proper estimation of the cluster boundaries is needed for discovering rare solutions. Indeed, as we show in a subsequent work, efficient approximate estimation of the droplets boundaries can be achieved, by similar methods as presented here, that can be applied to arbitrary graphs without having any detailed knowledge of the ground or low excitation manifolds~\cite{mohseni_nonequilibrium_2021}.

\begin{figure}[t]
\begin{centering}
\includegraphics[width=\columnwidth]{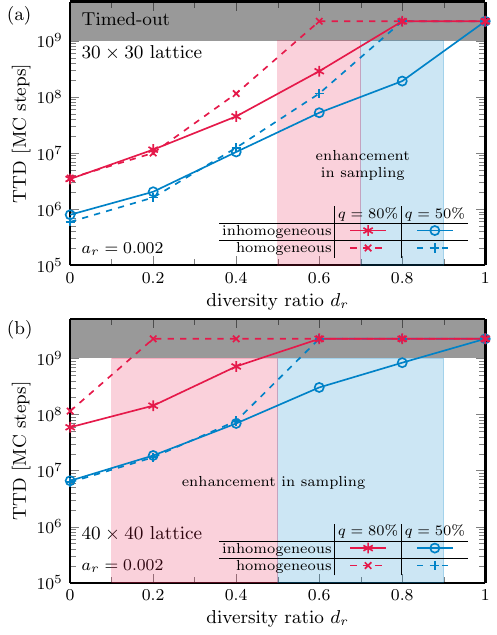}
\par\end{centering}
\caption{{\bf Time-to-diversity-ratio: homogeneous quench vs. a portfolio of inhomogeneous protocols in 2D disordered Ising model.} A portfolio of inhomogeneous fronts allows to substantially reduce TTD required to reach intermediate and large values of the diversity ratio $d_r$. It opens a way to sample from attraction basins that otherwise are inaccessible in reasonable times by the standard homogeneous driving. A separation (indicated with blue and red regions for $50\%$ and $80\%$ quantiles, respectively) grows with increasing system size, where we show a lattice of  $30\times30$ variables in panel (a) and $40\times40$ in panel (b).
We plot the median TTD ($q=50\%$), as well as the results for harder instances at $q=80 \%$ quantile (of 100 instances), where the separation looks more pronounced.
\label{fig:4}}
\end{figure}

\begin{figure}[!t]
\begin{centering}
\includegraphics[width=\columnwidth]{{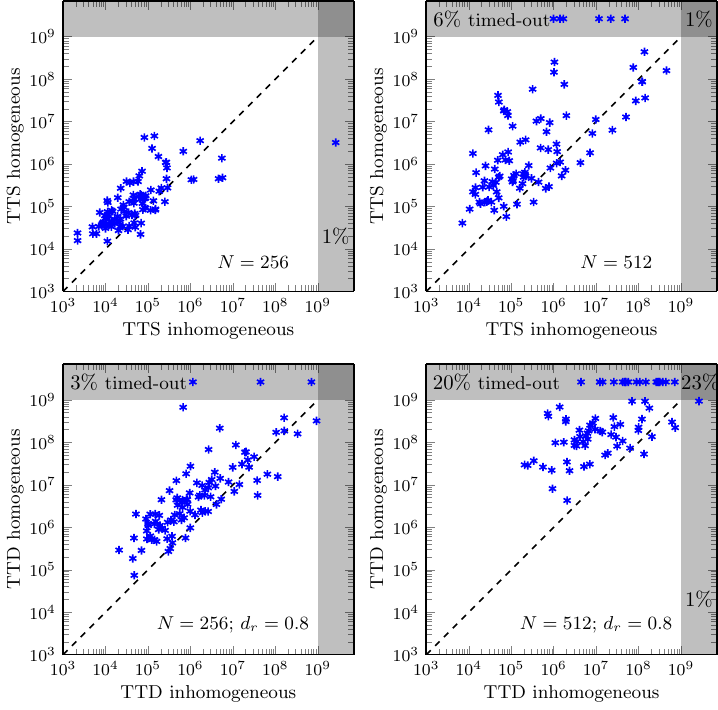}}
\par\end{centering}
\caption{
{\bf Comparison of homogeneous and inhomogeneous driving strategies in quasi-1D setup.} The scattered plots collect TTS and TTD for 100 disorder instances overviewed in Fig.~\ref{fig:3}(b). The first row shows TTS for a quasi-1D setup for $a_r=0.005$, and the second row shows the corresponding data for TTD to $d_r=0.8$ at normalized radius $R=1/8$. The gray bars indicate the timed-out instances, with digits giving the percentage of such instances (separately, for homogeneous protocol only in the top bar, for the portfolio of inhomogeneous quenches only in the right bar, and simultaneously for both strategies at the intersection of two bars). Of particular interest are the selected hard (timed-out) instances that become unfrozen by inhomogeneous strategy. Each set of instances and protocol is optimized over annealing times.
~\label{fig:s1}}
\end{figure}

\begin{figure}[!t]
\begin{centering}
\includegraphics[width=\columnwidth]{{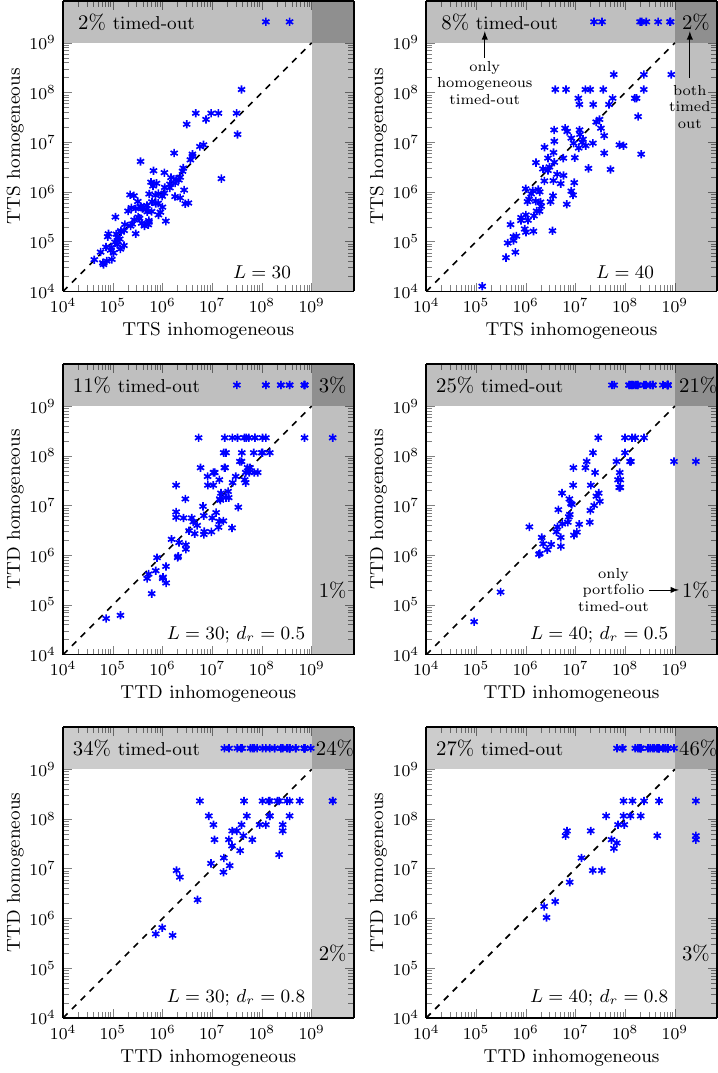}}
\par\end{centering}
\caption{
{\bf Comparison of homogeneous and inhomogeneous driving strategies for 2D disordered Ising model.} The data corresponds to the protocols studied in Fig.~\ref{fig:4}. The first row shows TTS for $a_r=0.002$ and $L=30$ (left column) and $40$ (right column), and the other rows focus on TTD to the targeted $d_r=0.5$ and $0.8$. Gray bars in the plots indicate that the solution has not been reached within the maximal allowed time. The inhomogeneous strategy allows to significantly reduce the number of time-outs for larger desired diversity ratios. Each system size and protocol is optimized over annealing time $t_a$, minimizing the number of time-outs, with the same $t_a$ used for all 100 instances.
~\label{fig:s2}}
\end{figure}

The advantage provided by inhomogeneous schedules, measured in terms of the required TTD for various targeted diversity ratios $d_r$ at fixed approximation ratio $a_r$, becomes significant with increasing system sizes where we see a growing separation between the performance of homogeneous and inhomogeneous schedules; see Fig.~\ref{fig:3}(b). The small-size effect appears to still be dominant for a setup with $N=128$ spins, where frequent repetitions of relatively fast quenches serve as the best strategy  (in that case, the optimal $t_a=2^8$ for the median instance, with the optimum moving towards smaller values of $t_a$ for harder instances).

The results for a 2D lattice are collected in Fig.~\ref{fig:4}. Here, the gains allowed by the algorithmic (multiple-fronts inhomogeneous) annealing schedule are even more noticeable. In particular, the standard homogeneous schedule is timing out with the increasing diversity ratio $d_r$, while the inhomogeneous one is still able to output some rare distinct low-energy configurations. According to that metric, for a $40\times 40$ spins system, we see achievable diversity $d_r$ enhanced by around $40\%$ of the maximum (both for a median and for harder instances at $80\%$ quantile). However, in the opposite limit of  $d_r \to 0$ in Fig.~\ref{fig:4}, one recovers the standard figure-of-merit of TTS, where the separation between the two protocols is negligible for median instances.

We collect further numerical evidence supporting such conclusions in the Appendix, where, in Figs.~\ref{fig:s1} and \ref{fig:s2}, we gather the scatter plots comparing the two driving strategies for all instances. Here, one can see that the fraction of hard-to-sample (timing out) instances can be reduced by more than 25$\%$ of the total, particularly in the limit of largest system sizes and targeted diversity ratios presented there.

We now comment on setting up multiple-front driving protocols simulated in our numerical experiments. Specifically, on dividing the lattice into clusters pictorially shown in Fig.~\ref{fig:2}. The tensor-network procedure of Sec.~\ref{sec:base} gives us reference diverse low-energy configurations seeding targeted basins of attraction. For quasi-1D systems in Fig.~\ref{fig:3}(b), we form the boundaries between clusters at positions corresponding to the boundaries of all droplets connecting those states with the ground state. We also absorb small clusters of few spins into neighboring ones---we only have clusters with sizes above 16 spins and an average cluster size of approximately 80 spins.

To simulate the quench dynamics in the quasi-1D setup, we employ the time-dependent variational principle (TDVP) for MPS~\cite{haegeman_time-dependent_2011,haegeman_unifying_2016} that projects the Schr\"odinger equation with nonlocal Hamiltonian into the tangent space of the MPS parametrization manifold. This allows one to integrate it by performing updates of individual MPS tensors. We consider 100 disorder instances for each system size and select the same critical fronts' slope $\alpha$ in all clusters. Finally, having the final state after the quench, we calculate the measurement probability for each low-energy configuration within the targeted $a_r$. It allows us to calculate the probability of obtaining a representative configuration from each basin of attraction. Those are used to estimate TTS and TTD as described in Appendix~\ref{sec:app1}. In this setup, we considered instances with the global reflection symmetry, and we merge basins differing by a global spin-flip transformation. The procedure is repeated for $\alpha = 0$, $1/128$, $1/64$, $1/32$, and a range of annealing times. We choose a sufficiently small approximation ratio of $a_r=0.0005$ to avoid benchmarking relatively easy problems for larger $a_r$ or facing vanishing diversity for too small $a_r$.

For the 2D setup, we estimate the probability of observing each targeted basin of attraction by employing QMC. Here, we follow an alternative approach to quasi-1D setup due to the algorithm's different nature. First, we consider a portfolio of clusters. Here, we start with large connected droplets between any pair of diverse low-energy configurations that seed targeted basins of attraction. We group those droplets into nonoverlapping sets (including their completion to the whole 2D lattice) forming a portfolio of clusters. Again, we do not form small connected clusters of a few spins here, absorbing them into neighboring clusters.
This results in a mean cluster size of approximately $300$ for $L=30$, and $600$ for $L=40$. For each QMC restart, we sample from that portfolio, and, for each cluster, randomly choose $\alpha$ from $0$, $1/50$, $1/20$, $1/10$, and $1/5$. We record the solutions for $a_r=0.002$ together with their distance from each of the targeted states. We count a solution as belonging to the closest basin of attraction, estimating the probability of reaching each basin. We perform simulations with $\alpha=0$ in all clusters  (i.e., standard homogeneous quench) as a reference.  In building the inhomogeneous portfolio, we can explicitly include a purely homogeneous one---we indeed used such a possibility here with a $1/5$ participation ratio. The seeds of the targeted attraction basins have been limited to be within $a_r=0.001$ due to numerical limitations in identifing them using our branch-and-bound baseline algorithm of Sec.~\ref{sec:base}. However, this limitation of attraction basins that are seeded by configurations below $a_r=0.001$, with counting a QMC solution as low-energy one when it is within $a_r=0.002$, removes the possibility to have the basins of attractions with only a few configurations within $a_r$, which would be artificially hard to sample.

\section{Conclusions}
We introduced a procedure to decompose the low-energy spectrum of a discrete optimization problem into likely independent clusters of solutions. This decomposition leads to a measure to quantify the diversity of independent, high-quality solutions within a given approximation ratio. We have examined this new measure on novel quantum annealing schedules. In particular, we have constructed algorithmic quantum annealing procedures by combining inhomogeneous driving strategy with efficient approximate tensor network contractions estimating distribution of topological defects characterizing low-energy domain walls. Such density and position of defects can be engineered by suitable choice of multiple inhomogeneous fronts driving the fluctuations in the system to minimize the residual energy of final states. We showed that such techniques can lead to sampling rare high quality solutions that are inaccessible by the off-the-shelf homogeneous strategies in the same timescales. In an accompanying work, the diversity measure introduced here was used as a new metric to experimentally quantify sampling power of quantum annealers against classical counterparts such as parallel tempering~\cite{zucca_diversity_2021}.

In addition to clustering the low-energy spectrum into likely independent {\emph types}, one could employ entropic measures of diversity on top of them, going beyond the number of classes $D$ that was employed in this work. Alternative strategies to estimate the boundaries of clusters can also be envisioned by using homogeneous annealing as a procedure to identify likely positions of defects, which can be used to set up the subsequent inhomogeneous driving protocols.

In related work, we show how one can adaptively learn nontrivial clusters of variables for generalized spin-glass Hamiltonians with k-local interactions, or arbitrary Boolean logic formulas in conjunctive normal forms, by efficiently estimating the geometry of the solutions in an instance-wise fashion~\cite{mohseni_nonequilibrium_2021}. This method leads to the construction of quantum-inspired nonlocal thermal annealing algorithms for sampling diverse sets of near-optimal solutions that are indeed sensitive to variations in the inputs, thus applicable to frozen regimes near computational phase transitions where OGP is expected to hold~\cite{mohseni_nonequilibrium_2021}.

\begin{acknowledgments}
We would like to thank Vasil Denchev and Vadim Smelyanskiy for useful discussions. M.M.R. acknowledges support by National Science Center, Poland, under Project No. 2020/38/E/ST3/00150, as well as receiving Google Faculty Research Award in 2017 and 2018.
\end{acknowledgments}
\renewcommand{\theequation}{A\arabic{equation}}
\setcounter{equation}{0}
\renewcommand{\thefigure}{A\arabic{figure}}
\setcounter{figure}{0}
\renewcommand{\thetable}{A\arabic{table}}
\setcounter{table}{0}
\begin{appendix}

\section{Calculation of TTS and TTD}
\label{sec:app1}
We calculate TTS in a standard way, namely, for $99\%$ success ratio, we employ the formula
\begin{equation}
{\rm TTS} = \frac{\log(0.01)}{\log(1-r) /t_s},
\label{eq:S1}
\end{equation}
where $r$ is the estimated success rate (probability of finding a solution within targeted $a_r$), and $t_s$ is the time of a single run (the number of sweeps in case of quantum Monte Carlo simulations of 2D systems, and annealing time $t_a$ for quantum annealing simulation in the quasi-1D setup).
Such a formula allows us to also average over the portfolio of solvers (e.g., combining runs with different front shapes $\alpha$ for the quasi-1D setup)
\begin{equation}
{\rm TTS} = \frac{\log(0.01)} {\overline{\log{(1-r^k)} / t^k_s}},
\label{eq:S2}
\end{equation}
where the overline indicates a mean over the set of solvers indexed with $k$,  that, e.g., might encode various values of $\alpha$ in our case. We use an algebraic mean for the quasi-1D simulations, meaning that all $\alpha$'s are equally weighted. In principle, this also allows one to consider situations where different protocols have different times of a single run, and, for algebraic mean, are selected at random with the same probability per unit of time.

To estimate TTD we have a success rate for observing a state belonging to each attraction basin $r_l$ with $l=1,2,\ldots,D$, which we order as $r_1>r_2>\ldots>r_D$. Analytical evaluation of TTD does not allow for a simple formula like Eq.~\eqref{eq:S1}. While one can resort to numerics, we approximate TTD by selecting $r=r_{l}$ for ceiling $l=\lceil D d_r \rceil$ and using formula in Eqs.~\eqref{eq:S1} and~\eqref{eq:S2}. This estimate becomes accurate in the case when there is a large separation between consecutive $r_k$ -- which is indeed the case for rare attraction basins and larger values of $d_r$. It overestimates the actual value for degenerate $r_l$, or in the limit of $d_r \to 0$ when the actual TTD converges to TTS. Still, we observe that using the approximate formula for TTD in the limit $d_r \to 0$ does not changes the results qualitatively.

We directly compare the TTS and TTD for homogeneous and inhomogeneous strategies for all considered disorder instances for a quasi-1D setup in Fig.~\ref{fig:s1} and for a 2D setup in Fig.~\ref{fig:s2}. For completeness, Table~\ref{tab:s1} includes an overview of total diversities for considered problems. We note that for the considered $a_r$ we do not see a clear correlation between diversity of a given instance and its hardness reflected by TTS and TTD.

The TTS (TTD) is optimized over annealing times for each panel and protocol. For instance, for $N=256$ in Fig.~\ref{fig:s1}, the globally optimal $t_a$ for homogeneous protocol TTS is $t_a=2^8$, and for a portfolio (including $\alpha=0$) $t_a=2^9$. It reflects a general trend that we observe in our examples, where optimal $t_a$ for inhomogeneous driving tends to be larger than for homogeneous driving. That is also the reason behind the appearance of a single extreme instance with TTS inhomogeneous timing-out (top left panel of Fig.~\ref{fig:s1}), as this instance gets trapped in local minimum for longer $t_a \ge 2^9$. It illustrates that forming a portfolio of annealing times might help solve such extreme points, however, at the cost of increasing typical TTS. Alternatively, we could also optimize $t_a$ for each solver and strategy included in the portfolio, which we have not done here for simplicity.
The optimal $t_a$ for TTD ($N=256$, $d_r=0.8$) are
$t_a=2^8$ for inhomogeneous and $t_a=2^7$ for homogeneous, which reflects another trend that we obverse, where increasing targeted diversity $d_r$ promotes using shorter $t_a$.

\section{Numerical details}
\label{sec:app2}
The reference low-energy subspace results discussed in Sec.~\ref{sec:base} have been obtained using the open-source implementation of Ref.~\cite{rams_spin-glass_2021} available at Ref.~\cite{TNAC4O_github}. Specifically, for the largest considered 2D examples of $N=1600$ spins, we obtain the final results using inverse temperature $\beta=5$, boundary-MPS bound dimension $\chi=64$, and $1024$ partial configurations kept during the search. The cut-off on the retained droplet size is $40$ and the energy cutoff corresponds to an approximation ratio of $0.001$. For corroboration, we combine the results of $4$ runs where the search (PEPS contraction) is performed from different edges of the square lattice and checked against different selections of algorithm parameters.

To simulate the sampling from quantum annealing quenches in Sec.~\ref{sec:quench}, in the quasi-1D setup we use TDVP for MPS~\cite{haegeman_unifying_2016} to integrate the real-time evolution. Due to the inhomogeneous and nontranslationally invariant nature of the setup, we combine one-site TDVP updates with only local application of more computationally expensive two-site updates. The latter is used to enlarge a given MPS bond dimension and is triggered based on Schmidt cutoff on a given MPS cut, which we set typically at $10^{-6}$, and the maximal bond dimension of up to 50 (which for selected points was checked for convergence against bond dimension 100). The time step $dt=1/8$ with $2$nd order integrator proves to be small enough due to relatively slow quenches. To avoid potential instability of TDVP applied to product states (the initial state at $g_i=1$), we start the dynamics in the ground state at $g_i=0.95$. The TTS and TTD results are optimized over a set of annealing times $t_a = 2^m$ with $m=4,5,6,\ldots,10$.

For annealing quenches in the 2D setup, we run QMC simulations at inverse temperature $\beta=24$. The number of QMC sweeps in a single repetition is optimized over the set $200$, $500$, $1000$, $2000$, \ldots, $100000$ (for $L=40$). The statistic is gathered over 25000 repetitions for each set of parameters.

\end{appendix}

%

\end{document}